\documentclass{PoS}
\title{Non-standard scaling laws of WIMP-nucleus interactions in WIMP direct detection}

\ShortTitle{Non-standard scaling laws of WIMP-nucleus interactions in WIMP direct detection}

\author{\speaker{Gaurav Tomar}\\
        Sogang University, Seoul\\
        E-mail: \email{tomar@sogang.ac.kr}}

\author{Sunghyun Kang, Stefano Scopel, Jong-Hyun Yoon\\
       Sogang University, Seoul\\
       E-mail: \email{francis735@naver.com, scopel@sogang.ac.kr, jyoon@sogang.ac.kr}}

\abstract{
Guided by the non-relativistic effective field theory of
interactions between Weakly Interacting Massive
Particles (WIMPs) of spin 1/2 and nuclei we study direct
detection exclusion plots for an example of non--standard 
spin--dependent interaction and compare it to the standard one. We
analyze an extensive list of 15 existing experiments including
the effects of momentum dependence and isospin violation. In our
analysis, we fixed the dark matter velocity distribution to a
Maxwellian.}

\FullConference{ICHEP2018, XXXIX International conference on High Energy Physics\\
		4-11 July 2018\\
		Seoul, Korea}

\begin{document}

\section{Introduction}
In the present paper we wish to extend the standard discussion
based on a Spin--Independent (SI) or Spin--Dependent (SD) WIMP--nucleus interaction using
the non--relativistic Effective Field Theory (EFT) approach of
Refs.\cite{haxton2, haxton1} and write the most general Hamiltonian density
describing the WIMP--nucleus interaction as,
\begin{eqnarray}
{\bf\mathcal{H}}({\bf{r}})&=& \sum_{\tau=0,1} \sum_{j=1}^{15} c_j^{\tau} \mathcal{O}_{j}({\bf{r}}) \, t^{\tau},
\label{eq:H}
\end{eqnarray}
where $c^0_j=\frac{1}{2}(c^p_j+c^n_j)$, $c^1_j=\frac{1}{2}(c^p_j-c^n_j)$, and
$\mathcal{O}_{j}$'s are the non-relativistic operators which can be found in~\cite{haxton2, haxton1}. In the following,
we consider a specific case of couplings $c_4$ and $c_6$ which 
correspond to the non--relativistic operators,
\begin{eqnarray}
  \mathcal{O}_4 &=& \vec{S}_\chi \cdot \vec{S}_N;\;\;\;\; \mathcal{O}_6=
  (\vec{S}_\chi \cdot {\vec{q} \over m_N}) (\vec{S}_N \cdot {\vec{q} \over m_N}).
\label{eq:ops}
\end{eqnarray}
The differential rate for the WIMP--nucleus scattering process is given
by:

\begin{equation}
\frac{d R_{\chi T}}{d E_R}(t)=\sum_T N_T\frac{\rho_{\mbox{\tiny WIMP}}}{m_{\mbox{\tiny WIMP}}}\int_{v_{min}}d^3 v_T f(\vec{v}_T,t) v_T \frac{d\sigma_T}{d E_R},
\label{eq:dr_de}
\end{equation}

\noindent where $\rho_{\mbox{\tiny WIMP}}$ is the local WIMP mass density in the
neighborhood of the Sun, $N_T$ the number of the nuclear targets of
species $T$ in the detector (the sum over $T$ applies in the case of
more than one nuclear isotope), while

\begin{equation}
\frac{d\sigma_T}{d E_R}=\frac{2 m_T}{4\pi v_T^2}\left [ \frac{1}{2 j_{\chi}+1} \frac{1}{2 j_{T}+1}|\mathcal{M}_T|^2 \right ],
\label{eq:dsigma_de}
\end{equation}

\noindent and, assuming that the nuclear interaction is the sum of the
interactions of the WIMPs with the individual nucleons in the nucleus:

\begin{equation}
  \frac{1}{2 j_{\chi}+1} \frac{1}{2 j_{T}+1}|\mathcal{M}|^2=
  \frac{4\pi}{2 j_{T}+1} \sum_{\tau=0,1}\sum_{\tau^{\prime}=0,1}\sum_{k} R_k^{\tau\tau^{\prime}}\left [c^{\tau}_j,(v^{\perp}_T)^2,\frac{q^2}{m_N^2}\right ] W_{T k}^{\tau\tau^{\prime}}(y).
\label{eq:squared_amplitude}
\end{equation}

\noindent In the above expression $j_{\chi}$ and $j_{T}$ are the WIMP
and the target nucleus spins, respectively, $q=|\vec{q}|$ while the
$R_k^{\tau\tau^{\prime}}$'s are WIMP response functions
which depend on the couplings $c^{\tau}_j$ as well as the transferred
momentum $\vec{q}$ and 
$(v^{\perp}_T)^2$. In equation (\ref{eq:squared_amplitude}) the
$W^{\tau\tau^{\prime}}_{T k}(y)$'s are nuclear response functions
and the index $k$ represents different effective nuclear operators,
which, crucially, under the assumption that the nuclear ground state
is an approximate eigenstate of $P$ and $CP$, can be at most eight:
following the notation in \cite{haxton2}, $k$=$M$,
$\Phi^{\prime\prime}$, $\Phi^{\prime\prime}M$,
$\tilde{\Phi}^{\prime}$, $\Sigma^{\prime\prime}$, $\Sigma^{\prime}$,
$\Delta$, $\Delta\Sigma^{\prime}$. The $W^{\tau\tau^{\prime}}_{T k}(y)$'s are function of $y\equiv (qb/2)^2$, where $b$ is the size of the nucleus. For the target nuclei $T$ used in
most direct detection experiments the functions
$W^{\tau\tau^{\prime}}_{T k}(y)$, calculated using nuclear shell
models, have been provided in Refs.~\cite{haxton2, catena}.
We will consider the possibility
that either of the couplings $c_{4}$ or $c_6$ dominates in the effective
Hamiltonian of Eq. (\ref{eq:H}) such that we can 
factorize a term $|c_j^p|^2$ by expressing it in terms of the {\it
  effective} WIMP--proton cross section\footnote{With the
  definition of Eq.(\ref{eq:conventional_sigma}) the WIMP--proton SI
  cross section is equal to $\sigma_p$, and the SD WIMP--proton cross
  section to 3/16 $\sigma_p$.},
\begin{equation}
\sigma_p=(c_j^p)^2\frac{\mu_{\chi{\cal N}}^2}{\pi},
  \label{eq:conventional_sigma}
\end{equation}
\noindent where $\mu_{\chi{\cal N}}$ is the WIMP--nucleon reduced mass and the ratio $r\equiv c_j^n/c_j^p$. 
It is important to point out that
except $M$ (SI interaction), $\Sigma^{\prime\prime}$ and $\Sigma^{\prime}$ (both
related to the standard spin--dependent interaction), all other nuclear response
functions vanish in the limit of $q\rightarrow$0,
and so allow to interpret $\sigma_p$ as a
long--distance point--like cross section. 
\section{Analysis}

The relative sensitivity of different
detectors is determined by the scaling law of the WIMP--nucleus cross
section with different targets, which is the focus of our analysis.
However the interaction terms in the Hamiltonian
of Eq.(\ref{eq:H}) lead to the expected rates that depend on the full set
of possible nuclear operators ($M$, $\Phi^{\prime\prime}$,
$\tilde{\Phi}^{\prime}$, $\Sigma^{\prime\prime}$, $\Sigma^{\prime}$,
$\Delta$) which imply different scaling laws of the WIMP--nucleus cross
section on different targets.  
\noindent We define the most stringent 90\% C.L. bound on the effective
WIMP--nucleon cross section $\sigma_{\cal N}$ as,
\begin{equation}
\sigma_{\cal N}=\max(\sigma_{p},\sigma_{n}),
  \label{eq:conventional_sigma_nucleon}
\end{equation}
\noindent which is a function of the WIMP mass $m_{\chi}$, and of the ratio
between the WIMP--neutron and the WIMP--proton couplings, $c_j^n/c_j^p$.

\begin{figure}
\begin{center}
\includegraphics[width=0.49\columnwidth]{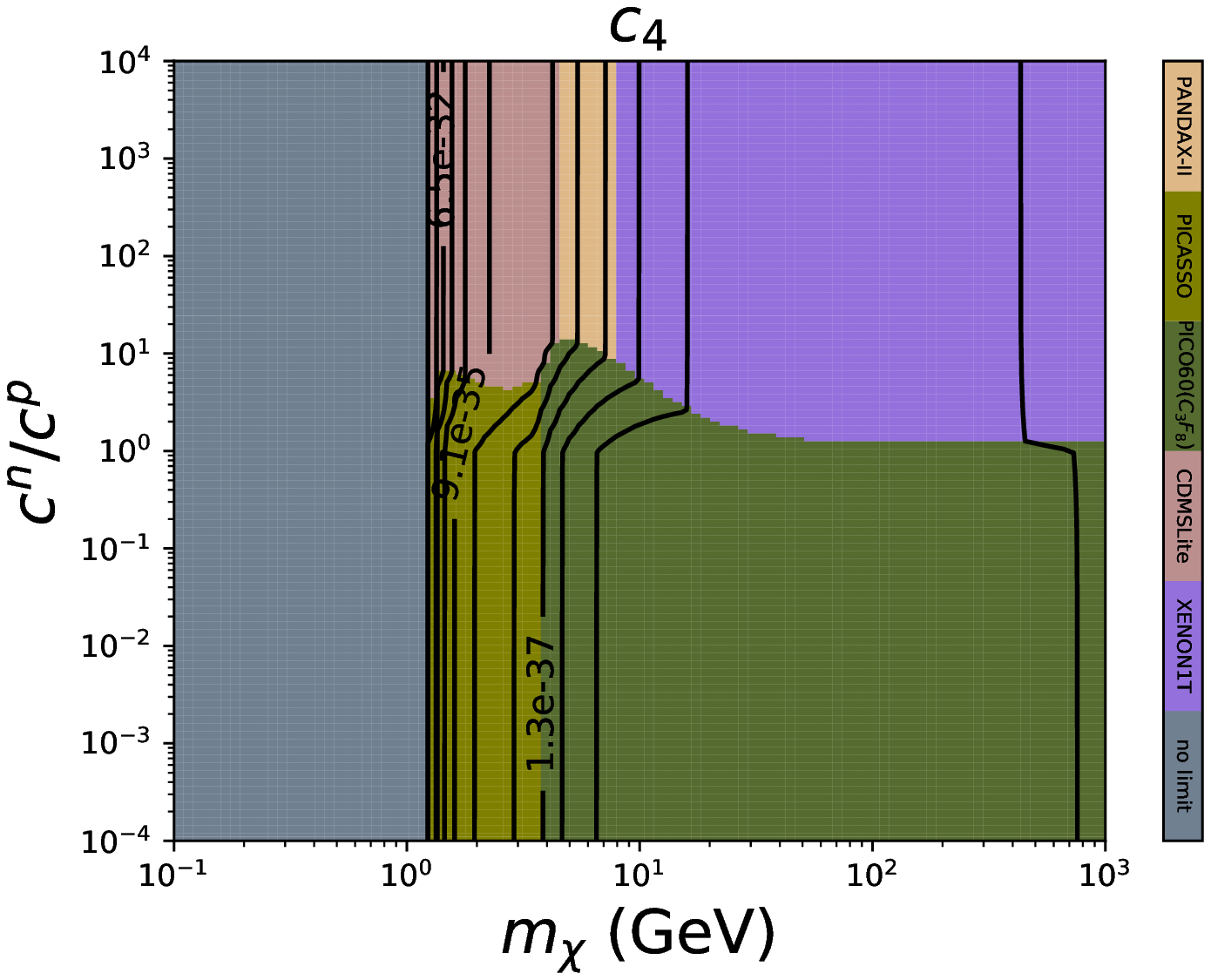}
\includegraphics[width=0.49\columnwidth]{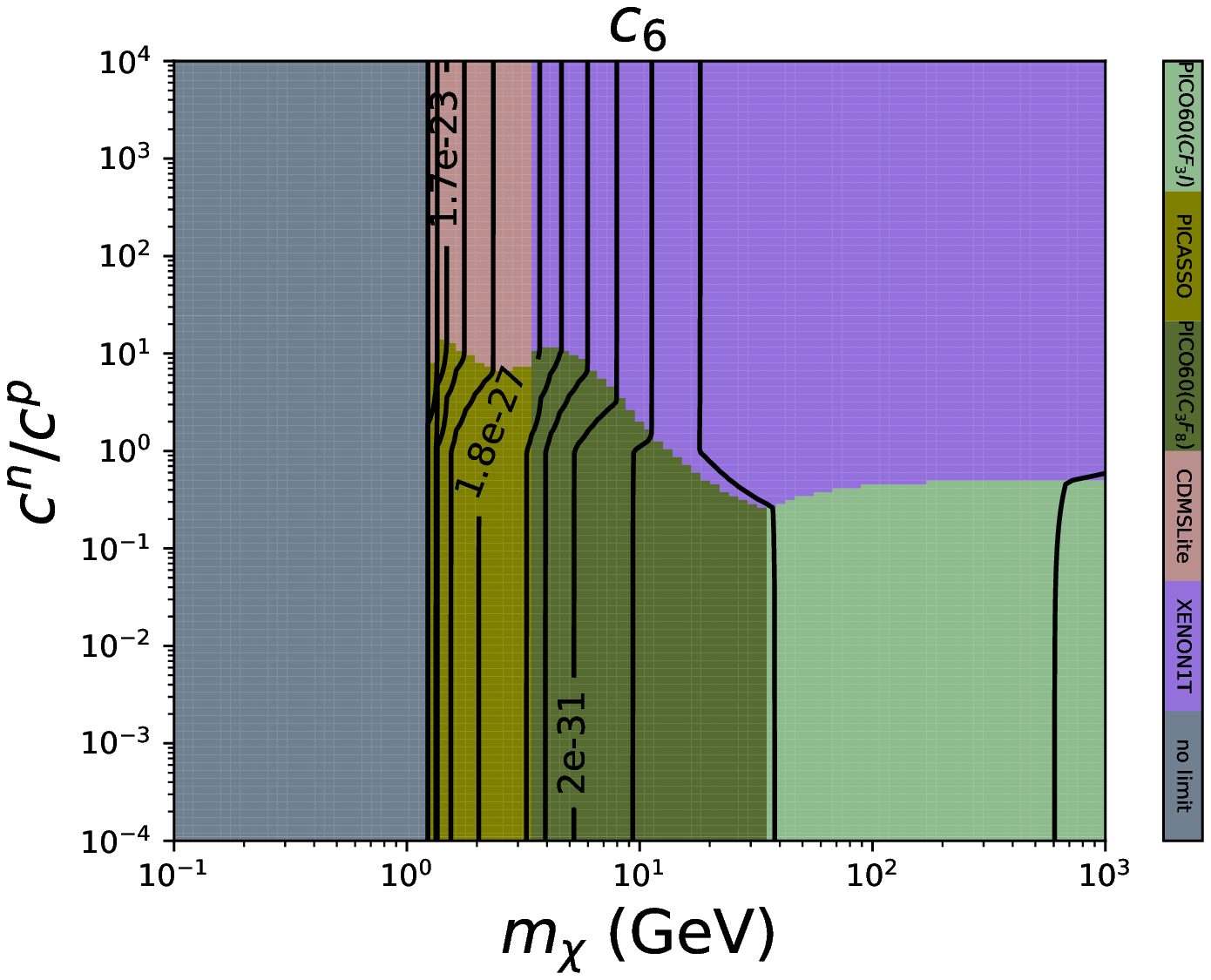}
\end{center}
\caption{Contour plots of the most stringent bound on the effective cross section $\sigma_{\cal N}$ (cm$^2$)
as a function of WIMP mass $m_\chi$ and of WIMP--neutron and WIMP--proton couplings ratio $c^n/c^p$
for operators $\mathcal{O}_4$ and $\mathcal{O}_6$. 
The experiments providing the most constraining bound are shown by different shadings and also indicated in the legend.}
\label{fig:c4_c6_plane}
\end{figure}

In Fig.~\ref{fig:c4_c6_plane}, we show the contour plots of $\sigma_{\cal N}$
(in cm$^2$) for the couplings $c_4$ and $c_6$.
While scanning the parameter space of $\sigma_{\cal N}$, the different regions of colors 
appear, which indicate the experiments providing the most constraining bound. 
The interactions terms $c_4$ and $c_6$ depend on the response functions 
$\Sigma^{\prime\prime}$
and/or $\Sigma^{\prime}$, that are related to the spin--dependent
coupling: basically, $\Sigma^{\prime\prime}$
corresponds to the coupling of the WIMP to the component of the
nucleon spin along the direction of the transferred momentum $\vec{q}$
while $\Sigma^{\prime}$ to that perpendicular to it, with
$W^{\tau\tau^{\prime}}_{\Sigma^{\prime}}(q^2)\simeq 2
W^{\tau\tau^{\prime}}_{\Sigma^{\prime\prime}}(q^2)$ in the limit of
$q^2\rightarrow 0$. Since inside nuclei the nucleons spins tend to
cancel each other, the contribution from even--numbered nucleons to the
response functions $\Sigma^{\prime\prime}$ and $\Sigma^{\prime}$ is
strongly suppressed. As a consequence, for such interactions
neutron--odd targets (such as xenon and germanium) are mostly
sensitive to the regime where $|c^n/c^p|\gtrsim$ 1 while proton--odd ones
(such as fluorine and iodine) mainly constrain the opposite case
$|c^n/c^p|\lesssim$ 1. This is reflected from the pattern of the shaded areas
of couplings $c_4$ and $c_6$ in Fig.~\ref{fig:c4_c6_plane}, where
for $m_{\chi}\gtrsim$ 1 GeV PICASSO (C$_4$F$_{10}$) and PICO (C$_3$F$_8$) bounds
(using proton--odd fluorine) are the most constraining limits for
$|c^n/c^p|\lesssim$ 1. Additionally, in the case of coupling $c_6$, 
PICO experiment with target CF$_3$I becomes equally competitive despite its relatively large
energy threshold ($E_R$=13.6 keV). This is because of the $q^4$ momentum dependence~(Eq.\ref{eq:ops})
that enhances the Iodine nuclear response function, while XENON1T and
PANDAX-II put the constraints for
$|c^n/c^p|\gtrsim$ 1. It is important to notice that at lower masses the constraint
is driven by CDMSlite, which is the experiment with a non--vanishing
spin target (germanium) with the lowest energy threshold. The low threshold 
experiments like DS50 and CRESST-II do not put any constraint in this regime
simply because argon and oxygen are the spinless nuclei. The similar results for the other couplings 
have been discussed in details in~\cite{Kang2018}, so we address the interested reader 
to that paper.
\section{Conclusion}
We have assumed a Maxwellian velocity distribution and explored in a systematic way the 
relative sensitivity of an extensive set of 15 existing DM direct
detection experiments to couplings $c_4$ and $c_6$
for the elastic scattering off nuclei of WIMPs of
spin 1/2.
We did our analysis in terms of two free
parameters viz. the WIMP mass and the ratio between the WIMP-neutron and
the WIMP-proton couplings $c^n/c^p$. The contour plots for each coupling is
provided in the $m_{\chi}$--$c^n/c^p$ plane of the most
stringent 90\% C.L. bound on the WIMP--nucleon cross section and
indicated with different shadings providing the most
constraining bound from the experiment. In the case of coupling $c_4$, the most constraining 
limit comes from PICASSO (C$_4$F$_{10}$) and PICO (C$_3$F$_8$) for 
$|c^n/c^p|\lesssim$ 1; XENON1T and PANDAX for $|c^n/c^p|\gtrsim$ 1. While for the coupling $c_6$,
PICO with Iodine target becomes equally competitive 
which is due to its $q^4$ momentum dependence.
This is evidence of the complementarity of
different target nuclei and/or different combinations of count--rates
and energy thresholds when the search of a DM particle is extended to
a wide range of possible interactions. The equivalent results for other 
couplings have been discussed in details in~~\cite{Kang2018}.
\bibliographystyle{JHEP}
\bibliography{test1}
\end{document}